\pdfoutput=1
\documentclass[letterpaper]{article} %
\usepackage[arxiv]{aaai25}  %
\usepackage{times}  %
\usepackage{helvet}  %
\usepackage{courier}  %
\usepackage[hyphens]{url}  %
\usepackage{graphicx} %
\usepackage{natbib}  %
\usepackage{caption} %
\usepackage{soul}
\usepackage{subcaption} %
\usepackage{algorithm}
\usepackage{algorithmic}

\usepackage{nth}
\usepackage{circledsteps}
\usepackage[nolist]{acronym}
\usepackage{booktabs}
\usepackage{xspace}
\usepackage{comment}
\usepackage{bm} %
\usepackage{todonotes}
\usepackage{arydshln} %
\usepackage{graphicx}
\usepackage{pifont} %
\usepackage{tikz}
\usepackage{soul} %
\usepackage{ifthen}
\usepackage{csquotes}
\usepackage[inline]{enumitem}
\usepackage{mathtools}
\usepackage{multicol}
\usepackage{cleveref} %

\usepackage{newfloat}
\usepackage{listings}
\usepackage{xcolor}  

\lstdefinestyle{mypython}{
    language=Python,
    basicstyle=\ttfamily\footnotesize,
    breaklines=true,
    frame=single,
    backgroundcolor=\color{white},
    keywordstyle=\color{blue},
    commentstyle=\color{green},
    stringstyle=\color{red},
    showstringspaces=false,
    tabsize=4,
    numbers=left,
    numberstyle=\tiny\color{gray},
}

\DeclareCaptionStyle{ruled}{labelfont=normalfont,labelsep=colon,strut=off} %
\floatstyle{ruled}
\newfloat{listing}{tb}{lst}{}
\floatname{listing}{Listing}
\title{Bootstrapping Social Networks: Lessons from Bluesky Starter Packs}

\author {
    Leonhard Balduf\textsuperscript{\rm 1},
    Saidu Sokoto\textsuperscript{\rm 2},
    Onur Ascigil\textsuperscript{\rm 3},
    Gareth Tyson\textsuperscript{\rm 4,5},
    Ignacio Castro\textsuperscript{\rm 5},
    Andrea Baronchelli\textsuperscript{\rm 2},
    George Pavlou\textsuperscript{\rm 6},
    Björn Scheuermann\textsuperscript{\rm 1},
    Michał Król\textsuperscript{\rm 2},
}
\affiliations {
    \textsuperscript{\rm 1}%
    TU Darmstadt\\
    \textsuperscript{\rm 2}%
    City St George's, University of London\\
    \textsuperscript{\rm 3}%
    Lancaster University\\
    \textsuperscript{\rm 4}Hong Kong University of Science and Technology (GZ)\\
    \textsuperscript{\rm 5}%
    Queen Mary University of London\\
    \textsuperscript{\rm 6}%
    University College London\\
    leonhard.balduf@tu-darmstadt.de,
    saidu.sokoto@city.ac.uk,
    o.ascigil@lancaster.ac.uk,
    gtyson@ust.hk,
    i.castro@qmul.ac.uk,
    andrea.baronchelli.1@city.ac.uk,
    g.pavlou@ucl.ac.uk,
    scheuermann@kom.tu-darmstadt.de,
    michal.krol@city.ac.uk
}

\newboolean{showcomments}
\setboolean{showcomments}{false}
\newcommand\important[1]{\todo[inline]{\textbf{Important:} #1}}
\newcommand\gareth[1]{\textbf{\textcolor{blue}{GT: #1}}}
\newcommand\leo[1]{\todo[color=orange,inline]{\textbf{Leo:} #1}}
\newcommand\michal[1]{\textbf{\textcolor{magenta}{Michal: #1}}}
\newcommand\onur[1]{\todo[color=red,inline]{\textbf{Onur:} #1}}
\newcommand\maciej[1]{\todo[color=violet,inline]{\textbf{Maciej:} #1}}
\newcommand\saidu[1]{\todo[color=yellow,inline]{\textbf{Saidu:} #1}}
\newcommand\ignacio[1]{\todo[color=cyan,inline]{\textbf{Ign:} #1}}
\newcommand{\pb}[1]{\paragraph{#1}}
\ifthenelse{\boolean{showcomments}} { }
{
	\renewcommand\important[1]{}
	\renewcommand\leo[1]{}
	\renewcommand\michal[1]{}
	\renewcommand\onur[1]{}
	\renewcommand\maciej[1]{}
	\renewcommand\saidu[1]{}
	\renewcommand\ignacio[1]{}
	\renewcommand\gareth[1]{}
}

\ifthenelse{\boolean{showcomments}}
{ \newcommand{\mynote}[3]{
		\protect\fbox{\bfseries\sffamily\scriptsize#1}
		{\small\textsf{\emph{\color{#3}{#2}}}}}}
{ \newcommand{\mynote}[3]{}}

\newcommand{\lb}[1]{\mynote{Leo}{#1}{blue}}

\newcommand{\eg}{\textit{e.g.}\@\xspace}

\newcommand{\ie}{\textit{i.e.}\@\xspace}

\newcommand\para[1]{\paragraph{#1}}

\newcommand{\one}{({\em i})\xspace}
\newcommand{\two}{({\em ii})\xspace}

\definecolor{verylightgray}{gray}{0.8}

\crefname{section}{Sec.}{Sec.}
\Crefname{section}{Section}{Sections}
\crefname{equation}{eq.}{eq.}
\crefname{figure}{Fig.}{Fig.s}
\Crefname{figure}{Figure}{Figures}

\floatstyle{plain}
\newfloat{lstfloat}{htbp}{lop}
\floatname{lstfloat}{Listing}
\crefalias{lstfloat}{listing}

\DeclarePairedDelimiter\len{\lvert}{\rvert}

\newcommand{\bsky}{Bluesky\xspace}

\DeclareMathOperator{\abs}{abs}

\usetikzlibrary{positioning}
\usetikzlibrary{graphs}
\usetikzlibrary{shapes.multipart}
\usetikzlibrary{shapes.misc}
\usetikzlibrary{matrix}
\usetikzlibrary{shapes}
\usetikzlibrary{arrows}
\usetikzlibrary{fit}
\usetikzlibrary{calc}
\usetikzlibrary{decorations.pathreplacing}

\tikzstyle{startstop} = [rectangle, rounded corners, minimum width=3cm, minimum height=1cm,text centered, draw=black, fill=red!30]
\tikzstyle{io} = [trapezium, trapezium left angle=70, trapezium right angle=110, minimum width=2.5cm, text width=2.5cm, minimum height=1cm, text centered, draw=black, fill=blue!30]
\tikzstyle{process} = [rectangle, minimum width=2.5cm, minimum height=1cm, text width=2.5cm, text centered, draw=black, fill=orange!30]
\tikzstyle{decision} = [diamond, aspect=3, minimum width=3cm, minimum height=1cm, text centered, draw=black, fill=green!30]
\tikzstyle{arrow} = [thick,->,>=stealth]
\tikzstyle{darrow} = [thick,<->,>=stealth]
\tikzstyle{entity_own} = [rectangle, minimum width=2.5cm, minimum height=1cm, text width=2.5cm, text centered, draw=black, fill=orange!30]
\tikzstyle{entity_foreign} = [rectangle, minimum width=2.5cm, minimum height=1cm, text width=2.5cm, text centered, draw=black, fill=gray!10]
\tikzstyle{functionality} = [rectangle, minimum width=2.5cm, minimum height=1cm, text width=2.5cm, text centered, draw=black, fill=white]
\tikzstyle{dataset} = [rectangle, minimum width=2.5cm, minimum height=1cm, text width=2.5cm, text centered, draw=black, fill=blue!10]
\tikzstyle{circlenode} = [circle, inner sep=0.05cm, draw=black, fill=gray!10, text centered]
\tikzstyle{ordernode} = [circle, thick, inner sep=0.02cm, draw=black, text centered]

\begin{acronym}[Derp]
    \acro{did}[DID]{Decentralized Identifier}
    \acro{nsfw}[NSFW]{Not Safe For Work}
    \acro{atp}[ATProto]{Authenticated Transfer Protocol}
    \acro{pds}[PDS]{Personal Data Server}
    \acroplural{pds}[PDSes]{Personal Data Servers}
    \acro{uri}[URI]{Uniform Resource Identifier}
    \acro{cbor}[CBOR]{Concise Binary Object Representation}
\end{acronym}

\begin{document}

\maketitle

\begin{abstract}
Microblogging is a crucial mode of online communication.
However, launching a new microblogging platform remains challenging, largely due to network effects.
This has resulted in entrenched (and undesirable) dominance by established players, such as X/Twitter.
To overcome these network effects, Bluesky, an emerging microblogging platform, introduced \emph{starter packs} --- curated lists of accounts that users can follow with a single click. 
We ask if starter packs have the potential to tackle the critical problem of social bootstrapping in new online social networks?
This paper is the first to address this question: we asses whether starter packs have been indeed helpful in supporting \bsky growth.
Our dataset includes
$25.05 \times 10^6$%
 users and
$335.42 \times 10^{3}$ starter packs with
$1.73 \times 10^{6}$ members, covering the entire lifecycle of \bsky.
We study the usage of these starter packs, their ability to drive network and activity growth, and their potential downsides. We also quantify the benefits of starter packs for members and creators on user visibility and activity while identifying potential challenges. 
By evaluating starter packs' effectiveness and limitations, we contribute to the broader discourse on platform growth strategies and competitive innovation in the social media landscape.
\end{abstract}

\section{Introduction}
Microblogging platforms have become integral to modern communication. Globally, over 4.5 billion people are active on social media, with microblogging platforms playing a critical role in this ecosystem~\cite{wearesocial2024}. For instance, studies show that over 70\% of users turn to social platforms to stay informed about breaking news and global events~\cite{pew2023}. These platforms also amplify diverse voices, enabling grassroots movements, social activism, and citizen journalism to thrive on a scale previously unattainable. 

However, launching a new social platform is extremely difficult. Users leaving an established platform are forced to leave familiar content, interfaces, and, most importantly, their social network. Of course, users could try to convince their social network to follow their migration. However, such attempts are rarely fully successful due to network effect~\cite{he2023flocking}. Worryingly, this exacerbates the dominance of established platforms, prevents innovation, and can constrain users' ability to migrate even if dissatisfaction with the platform is widespread~\cite{mekacher2024koo}.

Innovation attempts in microblogging are on the rise though. From blockchain-based microblogging such as Memo.cash~\cite{zuo2024understanding,10.1145/3543507.3583510}) to decentralisation attempts like Mastodon~\cite{he2023flocking}. 
Even large social networks such as Facebook have tried to innovate in this space~\cite{zhang2024emergence}.
Bluesky is part of this wave of innovation in microblogging. 
In 2022, Bluesky deployed a new microblogging service.
\bsky resembles Twitter/X: Users can follow each other and share short posts, including images and videos. %
A key innovation of \bsky is decomposing and opening the key functions of a social microblogging platform into sub-components that can be provided by stakeholders other than Bluesky~\cite{kleppmann2024bluesky}. 
The approach piloted by \bsky has enjoyed a large user adoption that has multiplied about 10 times its user base in a short period of time~\cite{balduf2024looking}:
over the last year, Bluesky's user base jumped
from %
$2.59 \times 10^6$%
 in January 2024
to %
$25.05 \times 10^6$%
 by the end of the year.
\bsky is now the largest new social platform, with over 26 million users.

However, \bsky still faces the  challenges in persuading users to migrate from incumbent competitors (\eg Twitter/X). To overcome this, \bsky introduced \emph{starter packs} in June 2024. 
Starter packs are curated lists of accounts that users can follow in a single click, enabling the rapid creation of a denser social network. Starter packs can be created by anybody and generally aim to (re)create new or existing communities. 

The rapid growth of \bsky gives credence to the ability of starter packs to mitigate the challenge of network bootstrapping and allow new users to quickly form social connections.
We believe that understanding the efficacy of starter packs is critical, both to understand the success of \bsky, as well as to identify the potential for other social networks facing similar challenges. If proven effective, starter packs could help disrupt future social networks by becoming a standard tool for onboarding and fostering early engagement. 

\paragraph{Research Questions.}
To address this we answer the following three research questions:

\begin{itemize} 
\item \textbf{RQ1:} How are starter packs used in \bsky, and to what extent are they employed?

\item \textbf{RQ2:} How effective are starter packs in driving network and activity growth? Do members of starter packs experience tangible social benefits? 

\item \textbf{RQ3:} How do users perceive the starter packs, and do they speak positively of them? Are there any downsides to introducing starter packs into the network?
\end{itemize}

First (\textbf{RQ1}), we collect \emph{all} starter packs, their changes, creators, members, and descriptions. We perform a temporal analysis, co-locate activity spikes with real-world events, and explore which communities use starter packs. 

We find that starter packs have experienced considerable uptake with 335,416 created over the 6 months since they were introduced. They are impactful, being responsible for up to 43\% of daily follow operations at their peak.
Yet, they include a relatively small number of users, with only 
$6.25 \, \%$%
 users being members of at least one starter pack. At the same time, the starter packs played an important role during large user influx spikes caused by various political events. We find that they are popular within artist, journalists, and academic communities.

Second (\textbf{RQ2}), we perform a temporal analysis of the \emph{follow} operations to estimate the number of new social graph edges created through starter packs. We then  quantify the benefits of being included or creating a starter pack using  Propensity Score Matching (PSM). We focus on increased visibility in the network (\eg a higher number of followers or received likes) and the activity of users (\eg a higher number of posts). We then use graph analysis to asses the macro-level impact on the overall social graph.

Our analysis reveals that becoming a member or a creator of a starter pack yields substantial benefits. 
Starter pack members receive up to 85\% more new followers and 70\% more likes than similar users not included in starter packs. 
Starter pack members also generate 60\% more posts and issue 71\% more likes.
This effect is even stronger for the starter pack creators reaching 117\% new followers and 100\% created posts increase.
On a macro-perspective, we notice a limited effect on the overall social graph.
Starter packs strengthen links between already existing communities rather than create new ones.
Furthermore, we find evidence that starter packs contribute to the \emph{rich get richer} effect,
increasing popularity inequalities in the system.

Third (\textbf{RQ3}), we extract all Bluesky posts discussing starter packs, perform sentiment analysis, and categorize those posts into the most commonly discussed topics.

We show that starter packs were mostly perceived positively by the community, with more than 10$\times$ more positive than negative posts.
At the same time, we notice multiple problems flagged by the users.
For instance, starter packs enable their creators to add any new member without asking for their permission.
This enables using popular and well-established accounts to promote malicious starter packs or use the feature as a tool of harassment.
Furthermore, we discover traces of a market where users pay to be included in a given starter pack. We make our code for identifying starter packs available to support future research.
\section{Background}

\para{Bluesky}
is a novel social network built on the \ac{atp}.
The system is decomposed into components that can be operated by the community.
We now introduce the relevant components for this study and refer interested readers to previous work~\cite{kleppmann2024bluesky, balduf2024looking} for a deeper analysis of the architecture and its critical components.

\para{User Data}
is stored in user-controlled repositories, each stored on a \ac{pds}.
Repositories provide signed and ordered lists of a user's public records.
Due to its open architecture, the repositories must be public and contain all the information required to operate the other components of the system.
Repositories store, for example, a user's posts, likes and follows, as well as other information such as the list of blocked users.

Repositories are updated via signed commits created by the user.
These commits include the creation of new records, as well as deletions or updates of existing records.
Commits are published via a publish/subscribe endpoint by the hosting \ac{pds}.

\para{Firehose}
is an aggregated publish/subscribe endpoint, which is subscribed to all federated \acp{pds} and re-publishes their commits as a single feed.

\para{Feeds}
are \bsky's algorithmically-driven content timeline creation mechanisms.
Each Custom Feed is generated by a Feed Generator, which curates posts to be included in the feed.
Feed Generators can be operated by \bsky or created by users.
When a user subscribes to a feed, the curated posts become available in the users' timeline, in the order stipulated by the Feed.
There is no limit on the number of feeds that a user can subscribe to.

\para{Starter Packs}
are an onboarding feature introduced to \bsky on %
2024-06-26%
.
A starter pack can be created by any user, without requiring any special privileges. Each starter pack consists of a list of up to 150 users and 3 feeds.
We refer to the users and feeds included in a starter pack as \emph{members}.
Starter packs also include a name, a description, and the name of their creator.
The contents of the starter pack are mutable and can be changed by its creator.

\bsky users can either \emph{(i)} use the \emph{follow all} option that automatically follows all the members  and subscribes to all the feeds in the starter pack; or \emph{(ii)} follow starter pack members individually.
\bsky enables users without a \bsky account to sign up via a starter pack. 
This triggers the regular sign-up process but also bootstraps the user's initial social network with a \emph{follow all} operation on the selected starter pack.

\section{Data and Methodology}\label{sec:methodology}

\subsection{\bsky{} Data Collection}
We collect a complete snapshot of \bsky by downloading data from all known \acp{pds} on 2025-01-01.
We then complete the data with Firehose updates, starting from 2024-06-10.
Combined, this enables us to recreate \bsky's complete state at any given time between 2024-06-10 and 2024-12-31.
We gather public data only (\eg no direct messages).

Our dataset contains the activity of all 
$25.05 \times 10^6$%
 \bsky users at the end of 2024.
It includes %
$1.55 \times 10^9$%
 follow relations in the social graph
$810.17 \times 10^6$%
 posts,
and %
$3.87 \times 10^9$%
 likes.
This includes all the $335{,}416$ starter packs created before 2025-01-01, their metadata, and modifications (\eg updates, deletions).
Notably, around $20\,\%$ of the created starter packs were deleted before 2025-01-01:
by the end of 2024 there were  $265{,}595$ starter packs.

\subsection{Identifying Starter Pack Usage}
\bsky does not explicitly record when someone uses a starter pack.
However, through manual analysis, we find that clicking on \emph{follow all} from a starter pack triggers a multi-follow operation, whereby multiple starter pack members are followed in rapid succession.\footnote{The multi-follow operation includes only starter pack members who were not previously followed by the user clicking on \emph{follow all}.}
This manifests as a sequence of repository commits containing up to $50$ follows each.
We leverage this to identify candidate starter pack users by selecting all whose repositories contain such multi-follow operations.
This represents a lower bound on the starter pack usage, as users can also manually follow specific starter pack members (instead of the whole starter pack) without triggering the multi-follow operations.

\paragraph{Mapping Multi-Follows to Starter Packs.}
We then attempt to assign the multi-follow operations to specific starter packs.
First, for each day, we reconstruct members in each starter pack from the Firehose data.
This is necessary because members can be added or deleted from starter packs over time.
As such, we consider the state of the starter pack at the time the multi-follow operation took place.

We then match each multi-follow operation $O_i$ to the starter packs $S_j$ containing the most similar users. Intuitively, if a user follows a large number of starter pack members in a single commit, we can be confident they did so via using that starter pack.
Thus, for every pair between multi-follow operation following members $M_{O_i}$ and a starter pack containing members $M_{S_j}$, we calculate a weighted set overlap score $s_{ij} \in [0,1]$:
$$
s_{ij}=f_{ij} \frac{\len{M_{O_i} \cap M_{S_j}}}{\min(\len{M_{O_i}}, \len{M_{S_j}})}
$$
where $f_{ij}$ denotes a factor to penalize large size differences:
$$
f_{ij}=1-\frac{\abs \left( \len{M_{O_i} \cap M_{S_j}} - \max(\len{M_{O_i}}, \len{M_{S_j}})\right)}{\max(\len{M_{O_i}}, \len{M_{S_j}})}
$$
We then assign each multi-follow operation $O_i$ to the starter pack $S_j$ with highest $s_{ij}$.
Note that $s_{ij}=1$ if the multi-followed members match perfectly the starter pack members $M_{O_i} = M_{S_j}$ and the user clicking on \emph{follow all} did not previously follow any starter pack members.

Out of the %
$5.69 \times 10^6$%
 multi-follow operations,
we find matches (\ie map the multi-follow to a specific starter pack) for
$99.88 \, \%$%
,
with a median best set overlap score of %
$0.75$%
.
%
%
%
%
%
%
%
%

%
%
%
%
%
%
%

\subsection{A Primer on Bluesky Growth}
\label{sec:basic_stats}

For context, we briefly present the current growth and trends of Bluesky. As a new social platform, this is intended to lay a foundation for the rest of the paper.

\begin{figure}[tb]
	\centering%
	\includegraphics[width=\linewidth]{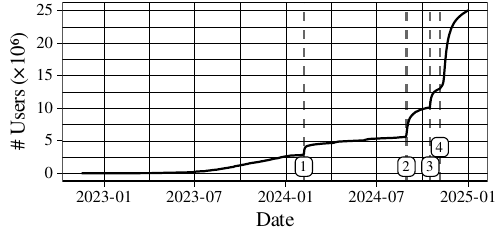}
	\vspace{-.7cm}%
	\caption{
		Number of registered Bluesky users.
	}
	\label{fig:user_count_annotated}
\end{figure}

\pb{User Growth.}
\Cref{fig:user_count_annotated} presents the number of registered Bluesky users over time, with event annotations. We observe substantial growth, especially since mid-2024. 
A number of events seem to have fueled \bsky adoption:
\Circled{1} Bluesky opening registrations to the public (\ie requiring no invites);
\Circled{2} Twitter/X banning in Brazil;
\Circled{3} Twitter/X's  controversial change making content visible to blocked users, and
\Circled{4} The 2024 US elections.

\begin{figure}[tb]
	\centering%
	\includegraphics[width=\linewidth]{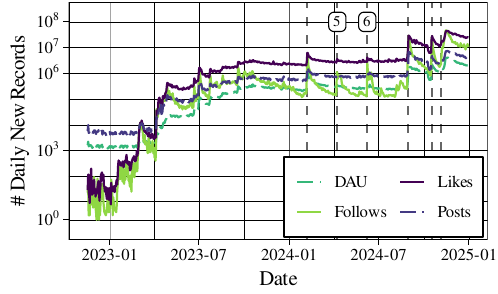}
	\vspace{-.7cm}%
	\caption{
		Number of Daily Active Users (DAU), likes, follows, and posts.
	}
	\label{fig:num_ops_per_day}
\end{figure}

\pb{Activity Growth.}
We now examine whether this user growth results in an increasing level of activity.
\Cref{fig:num_ops_per_day} plots the number of new follows, posts, and likes per day.
We confirm notable growth aligned with the influx of users.
\Circled{5} and \Circled{6} mark migrations and surges in activity from users 
from Brazil\footnote{\url{https://bsky.app/profile/bsky.app/post/3kufde3xvol2j}}
and Indonesia,\footnote{\url{https://bsky.app/profile/bsky.app/post/3kpllkdgtqu2v}}
both due to announced changes to Twitter/X, respectively.
That said, since early 2024, the number of posts has not grown linearly with the number of users.
Indeed, in December of 2024, there were an average of
$2.42 \times 10^6$%
 daily active users, representing just
$10.09 \, \%$%
 of daily registered users.
This suggests a large number of more experimental users, who are yet to actively engage in the platform.
The most common user activity is liking with %
$3.87 \times 10^9$%
 likes by the end of 2024,
compared to %
$1.55 \times 10^9$%
 follow operations and %
$810.17 \times 10^6$%
 posts.

\section{Measuring Starter Pack Use (RQ1)}
\label{sec:rq1}

Before assessing the impact of starter packs, we inspect the usage trends since \bsky introduced them in June 2024.

\begin{figure}[!ht]
	\centering%
	\includegraphics[width=\linewidth]{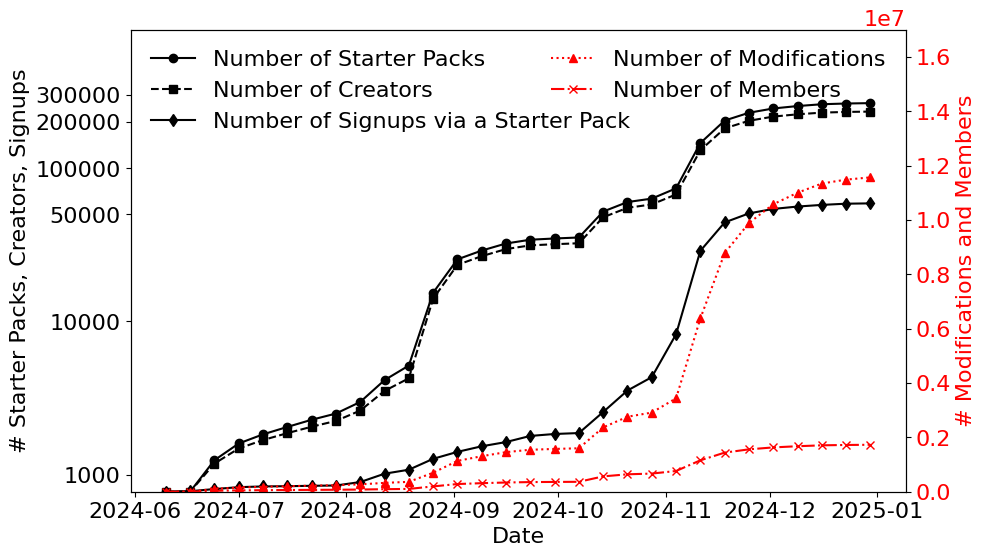}
	\vspace{-.7cm}%
	\caption{Evolution of the number of starter packs, their members, modifications, and creators over time.}	\label{fig:starter_growth}
\end{figure}

\pb{Starter Pack Growth}
Figure~\ref{fig:starter_growth} presents a time series of the number of starter packs released over time. We observe considerable uptake, with $265{,}595$ starter packs as of 2025-01-01. 
Recall, a total of $335{,}416$ starter packs have been created throughout the entire lifespan of \bsky --- this smaller number reveals that $69{,}821$ have since been deleted.
This uptake grew particularly after the first week of November 2024 (following the US elections), with a growth of 74\% from then to January 2025.
The number of creators is noticeably smaller than the number of starter packs, confirming that a subset of users create multiple ones. Indeed, 13.8\% of creators have two or more starter packs, and 3.4\% have more than three. We also find that starter packs are  actively maintained.
Recall that the creators can modify their starter packs --- 99.8\% of starter packs have at least one update, and 30.5\% have over 50. This suggests considerable investment by their creators and an active and evolving community.
By the end of 2024,
$238.51 \times 10^3$%
(%
$0.95 \, \%$%
 of all) users had created at least one starter pack, %
$1.56 \times 10^6$%
(%
$6.25 \, \%$%
) users were members of at least one starter pack,
and %
$1.1 \times 10^6$%
(%
$4.37 \, \%$%
) users had employed the \emph{follow all} operation on a starter pack.
This indicates that the new feature was used by a relatively small portion of users. We note again that this is a lower bound on starter pack-mediated follows as we can only confidently detect  bulk follow operations (rather than users who perhaps only follow one or two people from a given starter pack). We observe that only around 60,000 users have taken advantage of the ``sign-up via starter pack'' feature to join Bluesky, a figure significantly smaller than the number of starter pack users overall.

\begin{figure}[!ht]
	\centering%
	\includegraphics[width=\linewidth]{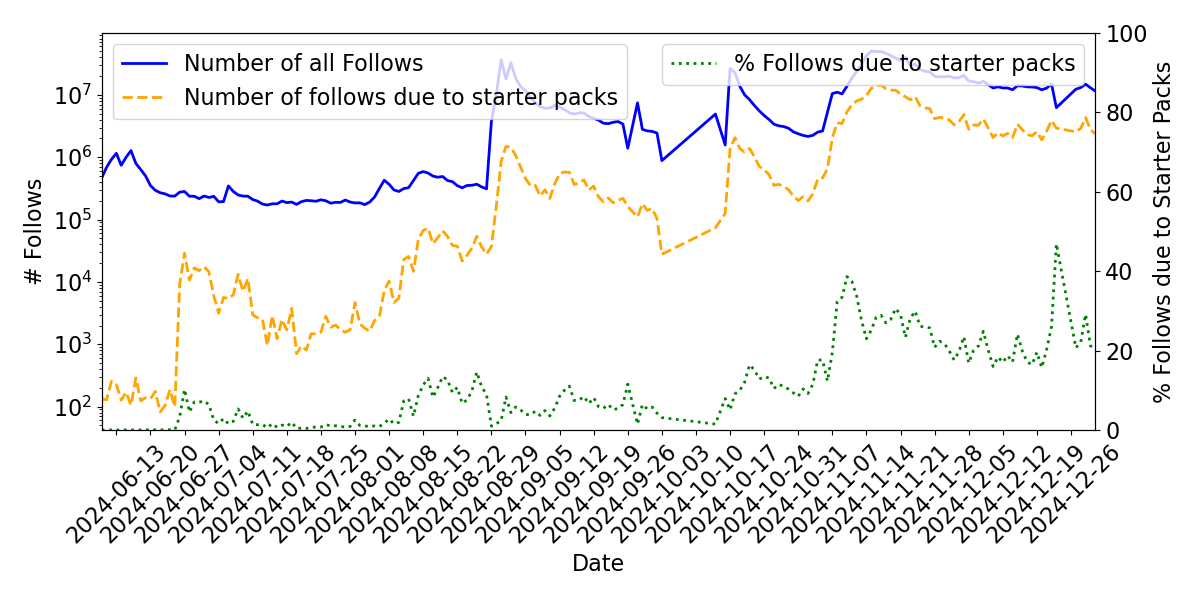}
	\vspace{-.7cm}%
	\caption{Daily count and percentages of all follow operations,   and follow operations due to starter packs.}
	\label{fig:follows_total_sp}
\end{figure}

\pb{Starter Pack Followers.}
We now estimate the number of follows created by starter packs using our matching of multi-follow operations resulting to starter packs, plotted as a timeseries in \Cref{fig:follows_total_sp}.
The starter-pack-induced follow operations closely match the system-wide trend.
Their impact on the social graph increases over time surpassing $40 \, \%$ of all the follow operations in December 2024.
We find that starter packs created a total of %
$308.57 \times 10^6$%
 unique edges in the follower graph.
This represents a remarkable $19.95 \, \%$ of all follow edges of the network,
indicating a large impact of starter packs on the overall social graph.
Follows resulting from starter packs are also long-lasting:
we observe that by the end of 2024,
$93.82 \, \%$%
 of them are still present.

\begin{figure}[htb]
	\centering%
	\includegraphics[width=\linewidth]{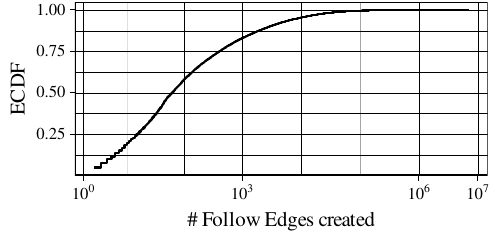}
	\vspace{-.7cm}%
	\caption{
		Distribution of follow edges created per starter pack.
	}
	\label{fig:starter_packs_num_follows_created_per_sp_ecdf}
\end{figure}

Figure~\ref{fig:starter_packs_num_follows_created_per_sp_ecdf} plots the number of followers created per starter pack.
As expected, the distribution is highly skewed, with the top $20\,\%$ of starter packs created 
$97.17 \, \%$%
of all starter-pack-induced follow edges.
The most popular starter pack (by the number of follow edges created)
with %
$7.01\,\text{M}$%
 follow edges created
lists \enquote{pro-democracy accounts}.
The following top 10 show a similar focus, with politics, journalism, and media as the main themes.

\begin{figure}[!ht]
	\centering%
	\includegraphics[width=\linewidth]{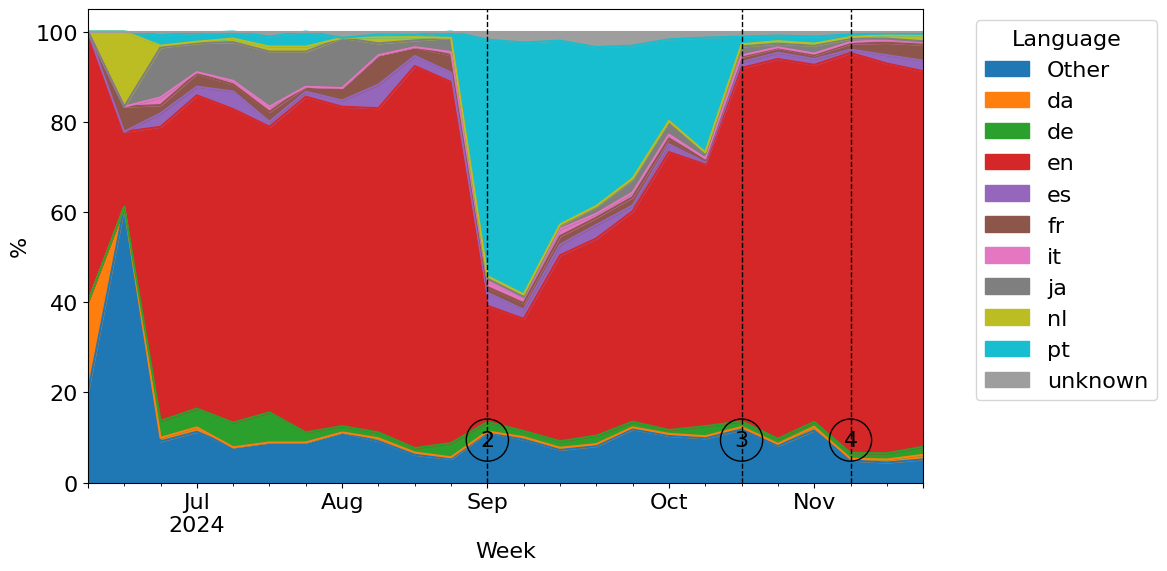}
	\vspace{-.7cm}%
	\caption{Language distribution in starter pack descriptions over time.}
\label{fig:language_distribution_of_starter_packs}
\end{figure}

\pb{Starter Pack Languages.}
We observe substantial use of starter packs across languages.
Figure~\ref{fig:language_distribution_of_starter_packs} shows the distribution of the top 10 languages in the starter pack descriptions, out of the 47 that we detect using \texttt{langdetect}.
We find that the prominence of the different language starter packs reflect underlying trends in new user arrivals (depicted previously in Figure~\ref{fig:user_count_annotated}).
For instance, the number of starter packs with a description in Portuguese spikes during the banning of Twitter/X in Brazil \Circled{2}, which corresponds to the period around the Twitter/X ban in Brazil (2024-08-30).
This dominance soon gives way to English, with two spikes that coincide with the \Circled{3} Twitter/X change in the visibility of blocked users (2024-10-17), 
and the \Circled{4} the US elections on 2024-11-05.

\begin{figure}[!htb]
\centering%
\includegraphics[width=\linewidth]{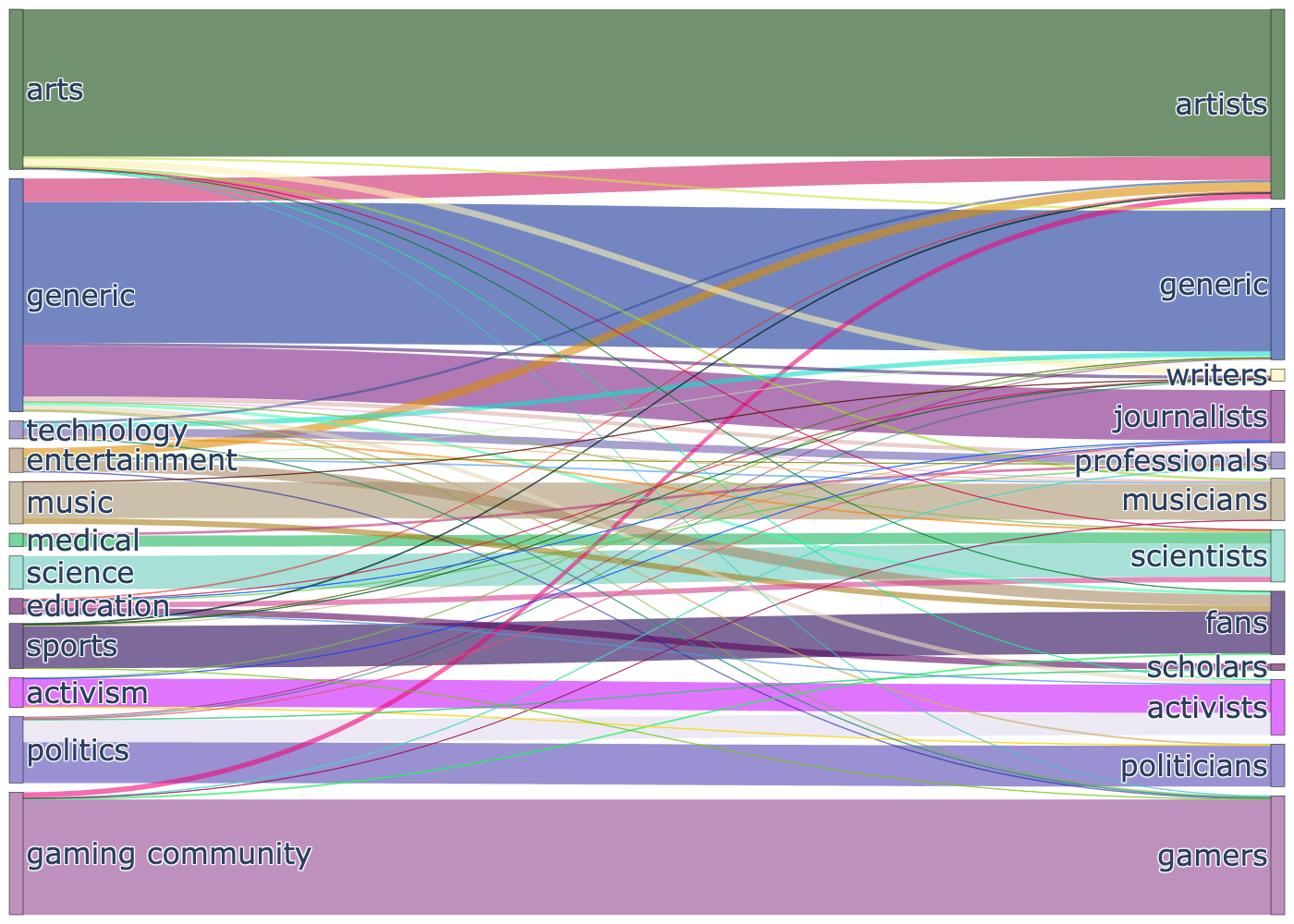}
\vspace{-.7cm}%
\caption{Sankey diagram with the theme of starter packs  (left) and the occupation or focus of their members (right).}
\label{fig:starter_pack_sankey_diagram}
\end{figure}

\para{Starter Pack Communities.}
To briefly explore the topics that these starter packs cover, we randomly sample and inspect 25\% of the 58,126 starter packs that include descriptions.
We employ the Large Language Model (LLM) Mistral~\cite{jiang2023mistral} (see prompts and details in the Appendix~\ref{app:post_classification_methodology}) for this classification.
We prompt the LLM to provide both the focus of the starter pack as well as the list of potential members, often listed in the pack description.
For instance a starter pack focused on climate  might be explicit about potential members being journalists, politicians, activists, or scientists.

We identify a large number of \enquote{generic} starter packs.
These are often the creator's personal selection (\eg \enquote{My favorite \bsky accounts}) without a specific focus indicated in the description.
Similar to related work~\cite{balduf2024looking}, we notice a well-developed art and gaming communities. 
We also observe starter packs focused on emerging communities such as politicians, journalists, or scientists.
This suggests that certain professional communities have been migrating to \bsky and are using starter packs to bootstrap their social graphs.

\section{Measuring Starter Pack Impact (RQ2)}
\label{sec:controversies}

Starter packs have seemingly helped \bsky to quickly bootstrap a large and active social network.
We now assess this by quantifying the impact of being included in or creating a starter pack.
This is non-trivial. Many factors may help a user gain followers, agnostic to their inclusion in a starter pack.
We address this challenge with Propensity Score Matching (PSM). PSM has proven effective in related tasks where controlling for several confounding variables is necessary~\cite{bhattacharjee2022does,valenzuela2014facebook,dos2015using}.
We then assess the macro-level impact of the starter packs using social graph analysis.

\subsection{Methodology}
\label{sec:psm}

\para{Propensity Score Matching (PSM)}
is a statistical technique that estimates the effect of a treatment or policy by accounting for the factors that predict treatment (\ie focusing on causation rather than correlation).

First, we divide \bsky{} accounts into treated and control groups.
We define two independent treatment indicators specifying whether: \emph{(i)} a user has been a member of at least one starter pack; and \emph{(ii)} a user has created at least one starter pack.

For each experiment, we calculate  success indicators measuring the increase in the \emph{(i)}  number of followers; \emph{(ii)} likes received; \emph{(iii)} issued likes; and \emph{(iv)} average daily posts.

Indicators \emph{(i)} and \emph{(ii)} (\ie followers and likes received) measure whether starter packs increase the visibility (i.e., popularity) in the network, while indicators \emph{(iii)} and \emph{(iv)} (\ie likes issued and average daily posts) focus on the increase in the activity of the users involved.
We measure each success indicator at time $t$ specifying the time of the first inclusion or creation of a starter pack (depending on the treatment group) and after 7-day intervals (\ie $t+7$, $t+14$, $t+21$, $t+28$).
To ensure that success indicators are calculated reliably at all intervals, we exclude accounts that were included in or created a starter pack after 2025-12-03 (\ie, 28 days before the end of our dataset), this discards $99{,}386$ ($0.4 \, \%$) accounts.

To measure the effectiveness of starter packs, we match treated accounts with the most similar accounts from the control (\ie untreated) group and compare their success indicators.
This requires selecting $t$ for the control group as well.
We thus set $t$ at a random time after their creation between 
2024-01-01 and 2024-12-03, then calculate the same success indicators at 7-day intervals (up to 28 days).
The robustness of the PSM is determined by the level of similarity between treated and non-treated accounts (\ie the higher, the better).
To maximize the matching quality, we choose three different $t$ for each account in the control group, effectively tripling its size.

\para{PSM Confounding Factors.}
PSM requires a formal set of confounding factors that may impact the dependent variables. 
To control for confounding factors and ensure a robust analysis, we select a comprehensive set of covariates that capture key characteristics of the accounts and their activity. All the covariates are calculated at $t$. \textbf{Number of followers} reflects the existing popularity of an account, which could independently influence subsequent follower growth~\cite{kwak2010twitter}.
\textbf{Account age (in days)} captures the amount of time an account has had to accumulate followers and activity, mitigating the effects of longer-established accounts naturally having larger followings~\cite{mislove2007measurement}. 
\textbf{Total posts} and \textbf{total likes} on posts measure engagement levels and content attractiveness, which are likely to affect follower acquisition~\cite{gilbert2009predicting}. 
\textbf{Total likes outgoing} represents the degree of interaction initiated by the account, capturing a behavioral aspect of network participation~\cite{golder2007rhythms}. \textbf{Followers-to-following ratio} serves as an indicator of influence and credibility, as accounts with a higher ratio may be perceived as more authoritative or desirable to follow \cite{cha2010measuring}. 
Last, \textbf{network size} is included to account for temporal variations in the overall network’s growth, ensuring that follower trends are not conflated with the increasing pool of users on the platform~\cite{ugander2011anatomy}. Together, these covariates provide a nuanced representation of the factors that may influence follower growth, allowing us to isolate the causal effect of inclusion in starter packs.

\para{Graph Analysis}
To compare the \bsky social graph, we use the directed graph $G$ of all follows at the end of 2024.
Within $G$, we mark edges created via starter pack multi-follows $S$.
For the analysis, we investigate properties of $G$ with and without $S$, \ie $G$ and $G\setminus S$.

To obtain $G$, we take all follows on 2024-12-31.%
We do not include accounts without any incoming or outgoing follow edges.
Furthermore, we remove self-loops and duplicate edges from $G$,
which do not occur naturally,
but can be added manually by tech-savvy users.
To  measure the relevance of the starter packs, we  create a second graph, $G\setminus S$, where we remove from $G$ any edges that were created via starter pack multi-follow operations.
We use NetworKit to analyze and contrast the resulting graphs~\cite{networkit}.
Note that, for the analysis of the follower graph at different points in time, we omit from $G$ any edges created after the selected date.

\subsection{Results}
We first quantify the impact of the starter packs for both their creators and members using PSM.

\begin{figure*}[ht!]
    \centering
    \begin{subfigure}[b]{0.48\textwidth}
        \centering
        \includegraphics[width=\textwidth]{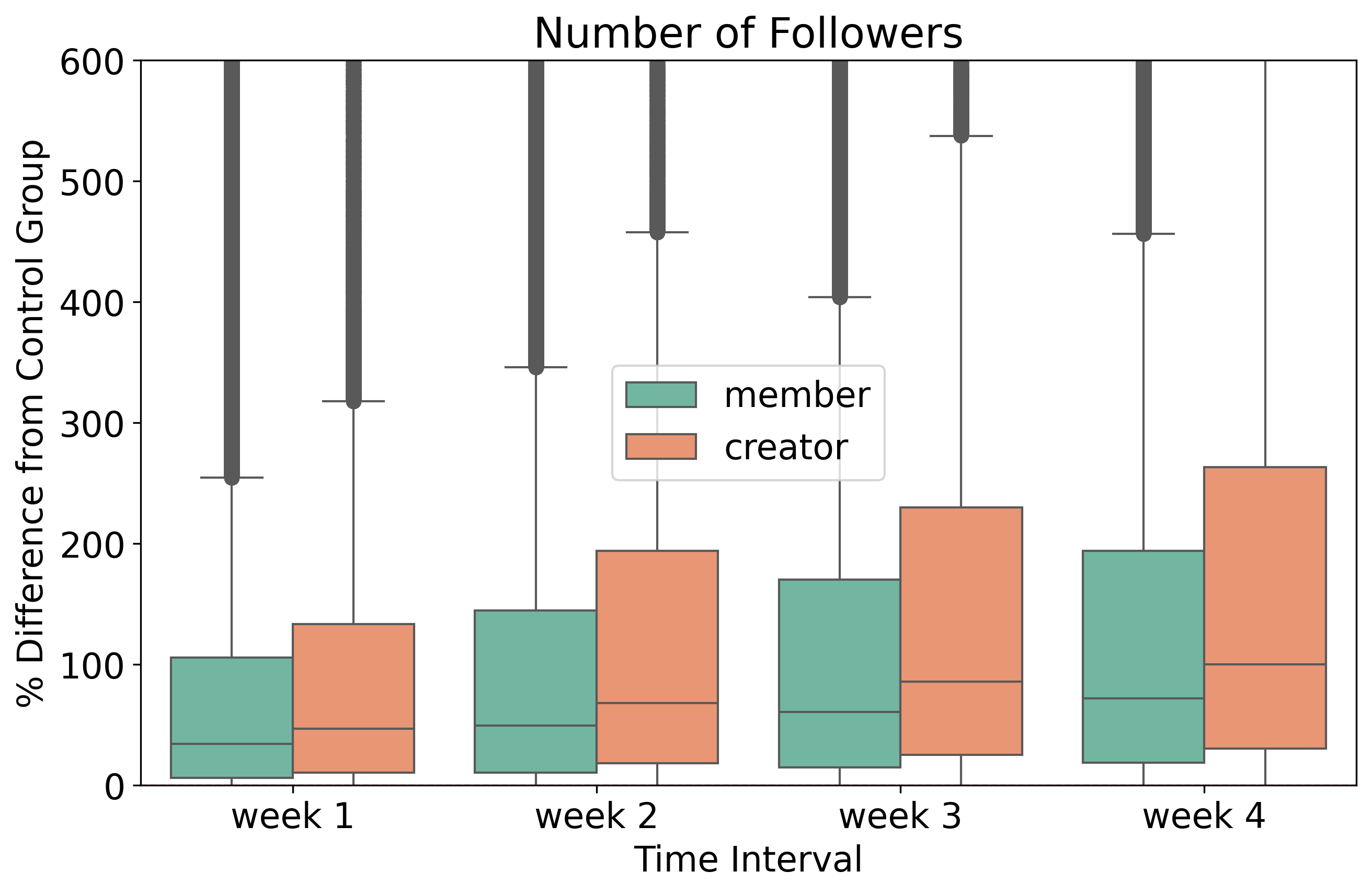}
        \caption{Followers number increase.}
        \label{fig:psm_followers}
    \end{subfigure}
    \hfill
    \begin{subfigure}[b]{0.48\textwidth}
        \centering
        \includegraphics[width=\textwidth]{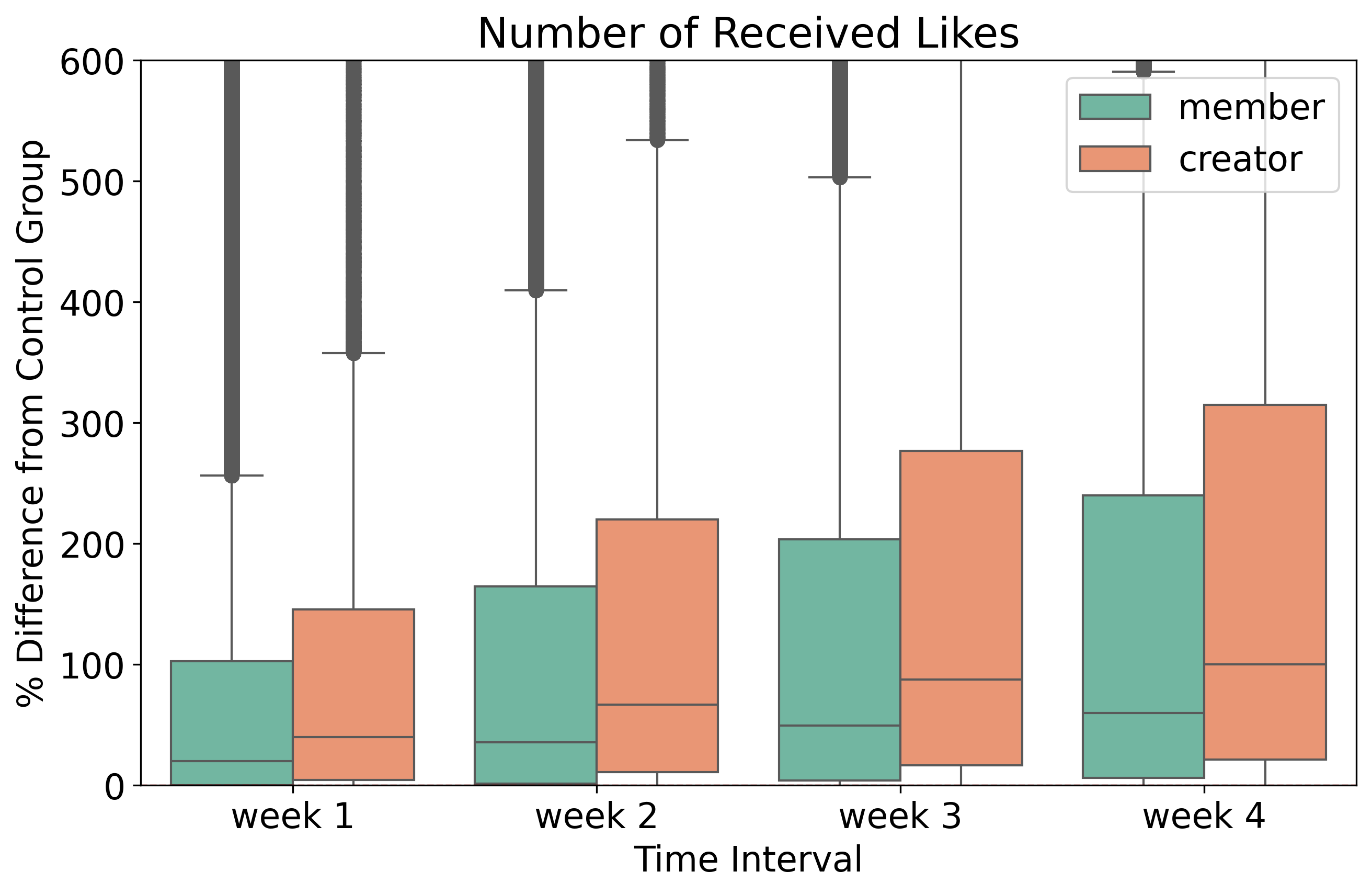}
        \caption{Received likes increase.}
        \label{fig:psm_received_likes}
    \end{subfigure}    
    \caption{Weekly visibility  increase for starter pack members and creators w.r.t. accounts in the control group (\ie neither creators nor members of an starter pack).}
    \label{fig:popularity_metrics}
\end{figure*}

\pb{Popularity Gains for Starter Pack Members.}
We investigate differences between accounts that are included in starter packs vs. those that are not.
We observe notable differences across all metrics.
We find that the popularity of a member grows after its inclusion in a starter pack. 
In the first week after its inclusion, the members receive on average 39\% more follow operations (\Cref{fig:psm_followers}).
This trend increases over time reaching 57\%, 71\% and 85\% after two, three, and four weeks, respectively.
We observe a similar trend for the number of likes received on published posts (\Cref{fig:psm_received_likes}).
The accounts included receive 23\%, 42\%, 51\%, and 70\% more likes in each of the four consecutive weeks.
This confirms that inclusion in a starter pack has a substantial positive impact on the visibility and popularity of the accounts included, and that the increase in popularity is not ephemeral.

\begin{figure*}[ht!]
    \centering
    \begin{subfigure}[b]{0.48\textwidth}
        \centering
        \includegraphics[width=\textwidth]{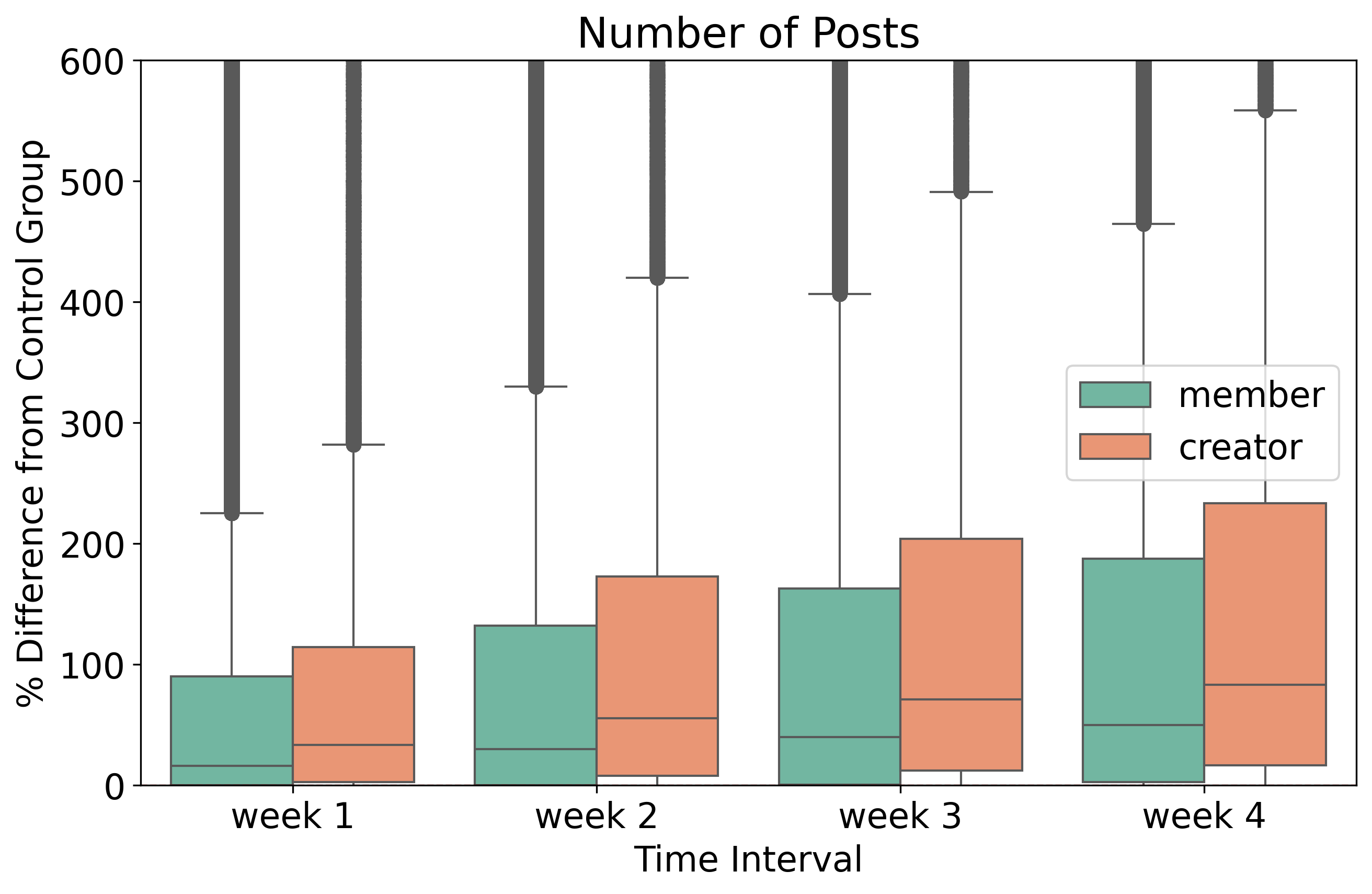}
        \caption{Number of written posts increase.}
        \label{fig:psm_posts}
    \end{subfigure}
    \hfill
    \begin{subfigure}[b]{0.48\textwidth}
        \centering
        \includegraphics[width=\textwidth]{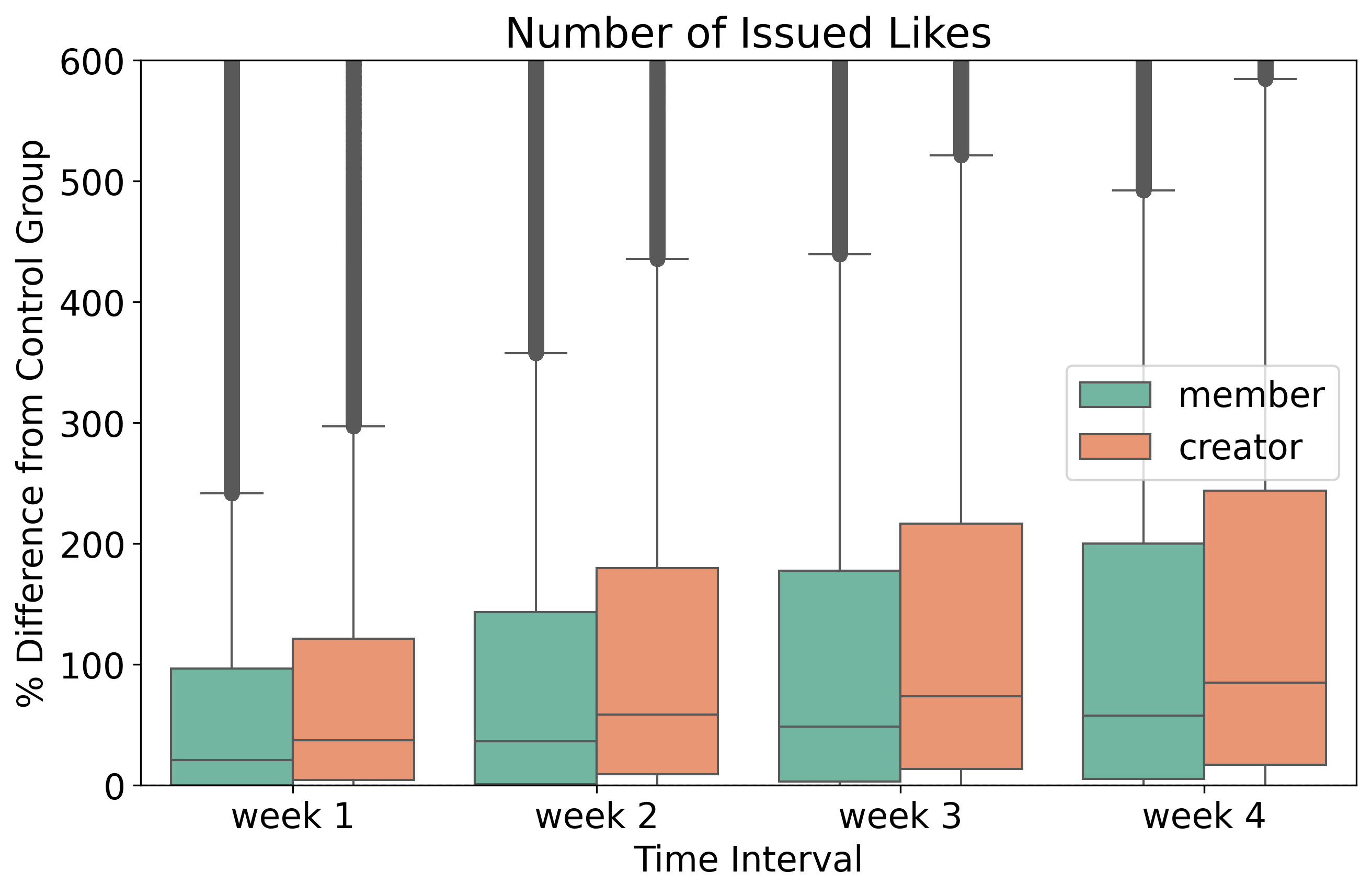}
        \caption{Number of issued likes likes increase.}
        \label{fig:psm_issued_likes}
    \end{subfigure}
    \caption{Weekly activity increase for starter pack members and creators w.r.t. accounts in the control group (\ie neither creators nor members of an starter pack).}
    \label{fig:activity_metrics}
\end{figure*}

\para{Activity Growth of Starter Pack Members.}
We then inspect the impact on the activity of the users who are included in a starter pack. We conjecture that inclusion in a starter pack may create a positive feedback loop that encourages greater activity (\eg posting).
Indeed, we observe a significant increase in the activity of the included accounts. 
The number of likes issued by members increases by 25\%, 45\%, 59\%, and 71\% in the 4 consecutive weeks (\Cref{fig:psm_issued_likes}).
At the same time, the number of posts created increases by 20\%, 36\%, 50\%, and 60\% (\Cref{fig:psm_posts}).
We hypothesize that the increase in popularity (and the related notifications) makes users more likely to visit the social network, and engage more widely.
This confirms that starter packs have a positive impact on encouraging engagement from both those who subscribe to them and those who are included in them.

\para{Benefits for Starter Pack Creators.}
Next, we investigate the benefits that accrue to the \emph{creator} of a starter pack. 
We conjecture that users who create starter packs may also gain increased popularity and activity. 

Indeed, in \Cref{fig:psm_followers} and \Cref{fig:psm_received_likes}, we observe an even stronger popularity increase with large increases in follows (51\%, 81\%, 100\%, and 117\%) and likes (46\%, 76\%, 100\%, and 115\%) across the four weeks. 
We further look at the activity of starter pack creators (\Cref{fig:psm_posts} and \Cref{fig:psm_issued_likes}).
Again, we observe a stronger positive impact compared to the inclusion in a starter pack.
This is manifested via a higher number of posts (40\%, 66\%, 87\%, and 100\%) and issued likes (45\%, 71\%, 90\%, and 102\%) across the four weeks.
This is likely caused by the starter pack creators being more engaged in the platform and feeling responsible for advertising the community from their starter pack.

It is also worth noting that $99.91\%$ of starter packs include the creators themselves. Naturally, this may mean that benefits largely stem from inclusion.
We, therefore, compare the benefits for developers of starter packs that include themselves vs. exclude themselves. We find that only 236 creators who do not include themselves still obtain some benefits.
At $t+28$, they experience an 8\% increase in received likes, a 22\% increase in new followers, issue 13\% more likes, and write 14\% more posts.
However, this is markedly less than creators who do include themselves.

\para{Sensitivity Analysis.} Our results rely on the assumption that the assignment of users to starter packs is unbiased. To evaluate the robustness of our conclusions to potential hidden biases, we conduct a sensitivity analysis based on the method outlined by Rosenbaum~\cite{rosenbaum2005sensitivity}.
This analysis quantifies how the likelihood of treatment within matched pairs could vary due to unmeasured confounding factors. 

Across all treatment subgroups (members and creators), user popularity and activity metrics, and time intervals, we observe a consistent sensitivity parameter $\Gamma = 3.0$. This indicates that within matched pairs, an individual’s likelihood of being treated could differ by a factor of up to 3, resulting in probabilities ranging from $1/(1+\Gamma) = 0.25$ to $\Gamma/(1+\Gamma) = 0.75$.
This does not alter our conclusion to reject the null hypothesis of no effect. These results demonstrate that the observed causal effects of starter packs on popularity and activity metrics \emph{are} robust to a substantial degree of unmeasured confounding.

\para{Graph Properties}
We next assess the starter packs macro-level impact on the \bsky social graph by comparing the follow relations graphs with ($G$) and without ($G \setminus S$) the starter pack edges.
Surprisingly, the removal of the $\approx 300 \, \text{M} $ starter-pack-induces edges
($\approx 20\, \%$ of all the edges) has little impact on the social graph from a macro-perspective. 
The number of strongly connected components increases from $8.25 \times 10^6$ to only $8.27 \times 10^6$ (\ie, a $0.002 \, \%$ increase).\footnote{
Note that strongly connected components are defined as maximal subsets for which a path exists between any two members.
As such, the addition of an edge can only decrease the number of components,
as it potentially collapses existing smaller components into one larger one.
}
The size of the largest strongly connected component decreases from $16.61 \times 10^6$ to $16.58 \times 10^6$ (\ie $0.002 \, \%$ decrease).
The average in- and out-degrees unsurprisingly do change, from $62$ to $50$, a decrease of $24 \, \%$.
This suggests that starter packs provide tighter connections within existing communities rather than creating inter-community connections and promoting connections across the entire system.
This raises the possibility of starter packs exacerbating potential echo chambers.

\begin{figure}[!h]
	\centering%
	\includegraphics[width=\linewidth]{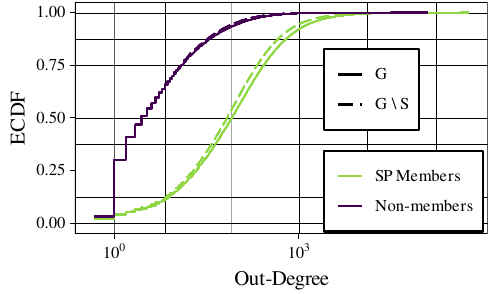}
	\vspace{-.7cm}%
	\caption{
		Out-degree distribution for the $G$ complete graph including all edges and the $G \setminus S$ graph without starter-pack induced edges, as well as for  members and non members of starter packs.
	}
	\label{fig:graph_analysis_out_degree_sp_vs_non_graph_comparison_ecdf}
\end{figure}

To better understand this phenomenon, we investigate the out-degree distribution (\ie number of followed accounts) for $G$ and $G \setminus S$ comparing starter pack members with the remaining users (\Cref{fig:graph_analysis_out_degree_sp_vs_non_graph_comparison_ecdf}).\footnote{
The in-degree changes affect only starter pack members, by definition, and are thus omitted here.
}
Interestingly, we find that removing starter pack induced edges almost exclusively affect  starter pack members, while the other users remain mostly unaffected.
This indicates that starter pack members are also the ones using starter packs, while non-members rarely use the starter pack \emph{follow all} operations.

\begin{figure}[!h]
	\centering%
	\includegraphics[width=\linewidth]{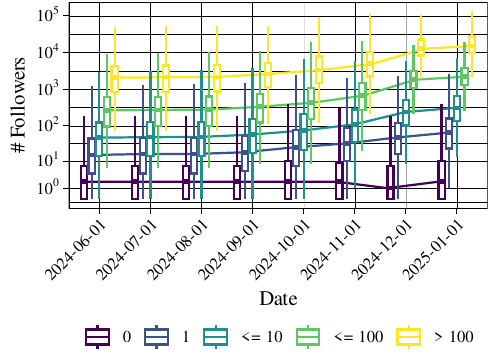}
	\vspace{-.7cm}%
	\caption{
		Evolution of the in-degree distribution  of  starter pack members depending on the number of starter packs they are included in at the end of 2024.
        Outliers omitted for visibility.
	}
	\label{fig:graph_analysis_eoy_sp_members_follow_evolution}
\end{figure}

The above points towards the fact that starter pack inclusion may be limited to already popular accounts in the network. 
To study this, \Cref{fig:graph_analysis_eoy_sp_members_follow_evolution} shows the number of followers over time for users who are included in different numbers of starter packs.
For each user, we calculate the number of starter packs they are included in on 2024-12-31.
Indeed, we observe that accounts included in many starter packs tend to be more popular at \emph{any} point in time, even before the introduction of starter packs --- note that the earliest  data point in the figure predates the introduction of starter packs on %
2024-06-26%
.
The graph shows the Matthew effect where popular accounts, are more likely to be included in starter packs increasing their popularity even further.
We observe the gap between more and less popular accounts widening over time.
We note that we can only detect the follow-all operations, and that users might instead cherry pick the specific accounts within a starter pack that they wish to follow.
These starter pack enabled follow operations might produce different results, although we posit that selecting specific individuals is more likely to increase the Matthew effect than bulk following all accounts in the the starter pack.

\section{Perceptions \& Downsides of Starter Packs (RQ3)}
\label{sec:rq3}
Finally, we briefly explore the users' perception of starter packs to understand whether they are appreciated or not by users.

\subsection{Methodology}
We extract a total of $363{,}999$ Bluesky posts ($0.06 \, \%$ of all posts since June 2024) containing term \enquote{starter pack} and analyze the sentiment of each post using the TextBlob Python library~\cite{textblob}.
TextBlob categorizes sentiment into three categories based on a polarity score: Positive (values $>0.1$), Neutral (values between $-0.1$ and $0.1$), or Negative (values $<-0.1$).

To better understand the content of the posts, we manually annotate 4,000 (1\% of all the posts mentioning starter packs) and identify the 11 most common themes expressed in positive, neutral, and negative posts. We then classify the remaining posts into those 11 themes using the Mistral LLM (see Appendix~\ref{app:negative_posts_classification} for the methodology). 
Note, we include an ``other'' category for all remaining posts that do not fall into one of the 11 themes.

\subsection{Results}

\begin{figure}[!ht]
	\centering%
	\includegraphics[width=\linewidth]{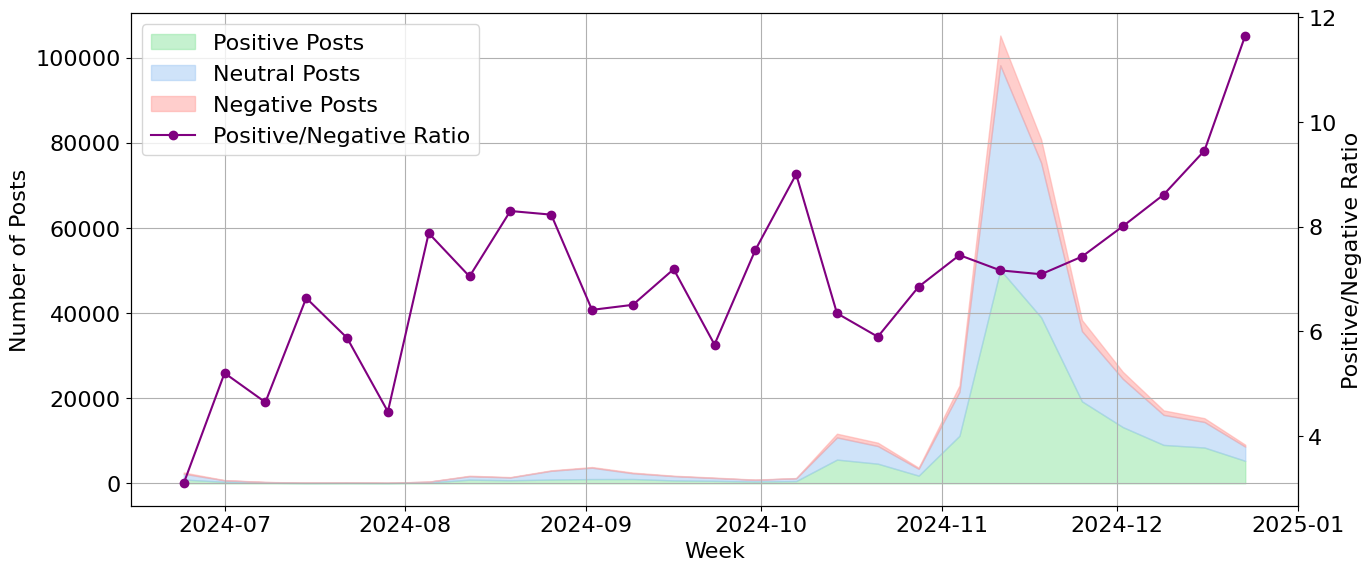}
	\vspace{-.7cm}%
	\caption{Distribution of the sentiment of posts discussing starter packs.}
	\label{fig:sentiment_analysis}
\end{figure}

\para{Sentiment Analysis.}
We identify a total of $176{,}648$ ($49\,\%$) positive and $164{,}225$ ($45\,\%$) neutral posts, yet only $24{,}127$ ($6\,\%$) negative posts.
\Cref{fig:sentiment_analysis} shows the volume of posts mentioning starter packs over time, classified into their corresponding sentiment.
Initially,  starter packs are rarely discussed.
Users start to discuss them broadly during the large influx of users to the \bsky network (\ie September--November, 2024). 
However, we note that starter pack discussion posts amount to only $\approx 0.06 \, \%$ of all posts since June 2024.
Confirm again that only a small number of \bsky users engage with the starter packs.

For the users who do discuss them,  starter packs are positively perceived, with $\approx 5\times$ more positive than negative comments in July (one month after the introduction of starter packs). 
The ratio increases over time with $\approx 10$ times more positive at the end of the year. 
As there were no major changes to how the starter packs work, our results suggest that the starter pack perception improves as the users start to use and learn more about them.

\begin{figure}[!ht]
	\centering%
	\includegraphics[width=\linewidth]{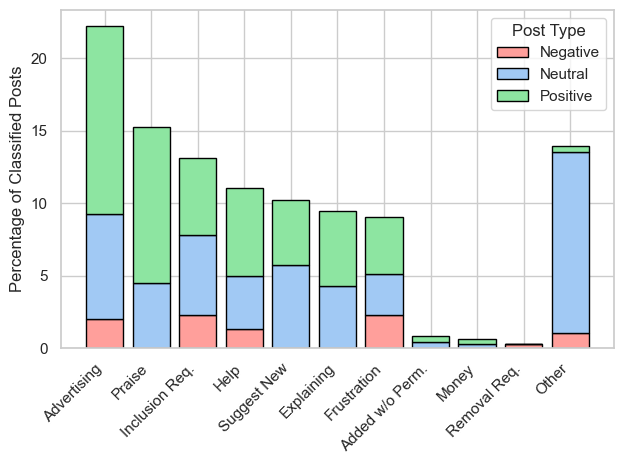}
	\vspace{-.7cm}%
	\caption{Distribution of posts mentioning starter packs by theme and sentiment%
	}
	\label{fig:visualizing_post_themes}
\end{figure}

\para{Theme analysis.}
To gain a deeper insight in the users' perceptions on starter packs, we also look into the themes discussed by the community.
\Cref{fig:visualizing_post_themes} plots a histogram of the 11 manually identified themes.
The most popular themes are advertising an existing starter pack or praising them. 
This is followed by users requesting to be included in an existing starter pack, suggesting a new one  or asking for help (\eg instruction on how to create a new starter pack).
Some users also express frustration with the current system and suggest new features to be added. 
Those posts focus mostly on the lack of a search feature making discovering new starter packs with
some users suggesting  notifications when an account is added to someone's starter pack.
Interestingly, we identified multiple instances of starter pack creators asking for money to include new members. 
This suggests the existence of an emerging market where popular creators can sell membership in their starter packs.
Multiple posts suggest that this exchange happens mostly over direct messages and its scale may be much larger than indicated by the public post analysis.

\para{User concerns.}
Finally, we manually analyze the posts classified under the \enquote{frustration} theme $1{,}699$ ($0.47\,\%$ of all posts containing starter packs)  to understand what potential concerns users have.
We find multiple users reporting that starter packs are used in a negative way. 
This includes using starter packs as lists of accounts to block or even harass. 
Such usage is usually inspired by political and ideological differences. 
Many users also dislike being added to starter packs without their consent, finding it stressful or invasive, especially when the resulting follower spike disrupts their usual interactions.
This includes popular and well-established accounts being added without their consent to malicious starter packs. It seems that creators do this to boost their starter pack credibility. 
We note that all these problems could be alleviated by requiring members to first agree to be added to a selected starter pack.

Multiple users also criticize the \emph{follow all} operation. These users perceive that it artificially inflates the social graph, leading to shallow interactions, and timeline pollution.
Finally, we observe an emotional strain reported by some starter pack members and creators.
Members worry that being added to starter packs may lead to followers expecting content misaligned with their usual posts, causing misunderstandings or negative feedback.
Creators sometimes report feeling pressured to include everyone --an impossible task given the cap in the maximum number of members of starter packs. They therefore fear backlash from excluded users.

\section{Related Work}

There are two core areas of related work: 
\one~social bootstrapping, often referred to as the cold start problem;
and
\two~social network migration.

\pb{Social Bootstrapping.}
There has been extensive work looking at the cold-start problem in social networks (aka ``social bootstrapping). 
Traditionally, this has been treated as recommendation system problem~\cite{yuan2023user}, whereby ``new'' friends must be recommended to the incoming user.
Early work focused on accelerating friend recommendations by quickly identifying other users with similar interests or friends-in-common~\cite{sahebi2011community}. However, these works do not consider user migrations from other platforms, instead, working on the assumption that users come afresh to the platform.
We argue that this is a missed opportunity as other studies have shown that prior social links have high predictive power in determining which newcomers will continue to engage with services~\cite{burke2009feed}.

Consequently, there have also been work looking at how new social networks can ``borrow'' links from older ones. 
Gong et al. \cite{gong2021cross,gong2018understanding} show that a user's existing social network (\eg on Facebook) can effectively be used to predict emergent links on a new social network (\eg Pinterest).
Zhong et al. \cite{zhong2014social} also study the benefits of migrating links from prior social networking platforms. They proposed a mechanism to copy social links for existing social graphs, called Link Bootstrapping Sampling. They show that borrowing certain links from existing user social networks does improve the robustness of the fledging social network.
Since these studies, various social networks have introduced such bootstrapping techniques. 
For example, Zhang et al. \cite{zhang2024emergence} explore how the creation of Threads (by Meta) benefits from the importation of links and users from Instagram.
Others have studied alternate applications of link transfers across social networks. For example, Venkatadri et al. \cite{venkatadri2016strengthening} use link transfers to establish trust in new social networks, which have not yet been bootstrapped. 
Bluesky is rather different, in that it does not migrate links directly from prior social networks. Instead, key people create starter packs that contain well-known people from the community. We show that this novel approach brings similar benefits to network bootstrapping

\pb{User Migrations.}
A small set of recent studies have investigated migration patterns between social networks, primarily of users from Twitter/X~\cite{bittermann2023academic}.
Rather than borrowing links, these primarily study cases where users entirely abandon the previous social network. 
He et al. \cite{he2023flocking} study the recent migration of users from Twitter/X to Mastodon. They show that the social network plays a major role, with users becoming more likely to migrate once their friends have migrated. Cava et al. \cite{cava2023drivers} found similar patterns, showing that a user's social network is a key factor in driving their migrations
Jeong et al. \cite{jeong2024exploring} also investigate the behavior of users who perform this migration.
We complement this work, by studying how such social networks are migrated on \bsky. Importantly, Mastodon lacks any concept of starter packs, forcing users to manually rebuild their network.
Importantly, our work differs in that we are not investigating \emph{why} users migrate to Bluesky. Instead, we focus on how starter packs simplify this migration and accelerate social bootstrapping.
\section{Conclusion \& Future Work}

This paper has studied the use of starter packs on Bluesky, exploring how it can help bootstrap a robust social network. 
We started by gathering all $335.42 \times 10^{3}$ starter packs in Bluesky, tracking their changes, creators, members, and descriptions (\textbf{RQ1}). Our temporal analysis identified activity spikes, particularly aligned with real-world events. Through this, we confirmed the central role of starter packs in facilitating organic community development.
This led us to examine the temporal patterns of followers on Bluesky (\textbf{RQ2}).
We found that being added to a starter pack \emph{does} brings significant benefits, with included users achieving far more followers than their counterparts.
We show that this further leads to increased activity, with such users sharing more posts. Despite this, we found that the effect of starter packs is mostly restricted to their members and it might reinforce the \textit{rich get richer} effect potentially helping to increase inequalities.
We then analyze posts discussing starter packs (\textbf{RQ3}). Although the majority of users speak positively about starter packs, our user perception analysis also revealed posts complaining about the effect of starter packs on the quality of their conversations and the dynamics around member inclusion.

By addressing the above research questions, we offer a foundation upon which others can study the influence of starter packs on social graph dynamics. We have a number of future lines of work. First, we are keen to expand our analysis over a more longitudinal time frame. In the several months we study, we already see noticeable temporal patterns. We are therefore interested in understanding how these evolve over longer periods, particularly outside of the large user influxes experienced in late-2024. Second, we argue that starter pack operators may increasingly play a prominent central role in Bluesky, impacting the formation of communities. This may lead to certain power imbalances, as well as lucrative business opportunities, \eg selling positions in a starter pack. We therefore wish to study this phenomena as it emerges. We are also keen to study if this creates potential echo chambers that could result in polarization.
We will make  code publicly available, enabling others to study starter packs and \bsky. 

\bibliography{refs}

\clearpage

\appendix
\section{Appendix: Methodology for Post Classification}
\label{app:post_classification_methodology}

Here we describe the methodology used to classify posts into predefined categories. The classification process involves a prompt-based approach using the Mistral language model. The analysis were performed on a Linux machine with 46 GB of RAM, 40 CPU cores (Intel Xeon Silver 4114), and an NVIDIA Quadro P4000 GPU.

\subsection{Starter Packs Classification Prompt}
The classification of starter packs and their participants is performed using the following prompt:

\vspace{0.5em} 
\begin{lstlisting}[basicstyle=\ttfamily\footnotesize, breaklines=true, frame=single]

Based solely on the provided name and description of the starter pack, classify it into a community/category.
Name: {name}
Description: {description}

Follow these instructions for the response:

1. Provide two classifications:
    - The first classification should represent the starter pack itself. If there is insufficient information or if the description is unclear, classify as "unknown".
    - The second classification should represent the participants (e.g., "artists," "musicians," "politicians," "activists," "scientists," "unknown"), as you deem appropriate.
      If there is insufficient information or if the description is unclear, classify as "unknown".

2. Each classification should be a category that represents the core idea of the community or its members.
    - Do not use two or more words (e.g., "sports or water sports").
    - Avoid ambiguity or overlapping terms. Select only the most appropriate classification based on the description and do not add any other details, just the classification.
    - If the description is unclear or if there is insufficient information, classify as "unknown".

3. Provide your response in the following format:
    1. Starter Pack Classification: [only a suitable classification for the starter pack or "unknown"]
    2. Participants Classification: [only a suitable classification for the participants (plural)]
\end{lstlisting}

\subsection{Posts Classification Prompt}
\label{app:negative_posts_classification}
The classification of posts is performed using the following prompt:
\vspace{0.5em}
\begin{lstlisting}[basicstyle=\ttfamily\footnotesize, breaklines=true, frame=single]
Classify the following post into one of these categories only.
Provide no additional text, explanation, or reasoning, just the category.

Categories:
    1: "Praising a Starter Pack or Starter Packs in General",
    2: "Explaining How the System Works or Reporting Starter Pack Experience",
    3: "Desire to Be Added to a Starter Pack",
    4: "Advertising a Starter Pack (including asking for members or inviting others to join)",
    5: "Expressing Frustration with the Current System (e.g., mass follow but zero engagement)",
    6: "Added Without Permission",
    7: "Suggesting Someone Create a New Starter Pack",
    8: "Asking for help (e.g., understanding how the system works or looking for a specific Starter Pack)",
    9: "Asking for money to include someone in a starter pack",
    10:"Asking to be removed from a starter pack",
    11: "Other"

Post: {post}

Your response should follow this exact format:
Category: [chosen category from the 9 options]

Guidelines:
- Do not add any text, parentheses, explanations, or reasoning.
- If unsure, select: "Other not covered by the above categories."
- Output only the category in the specified format.
\end{lstlisting}
\vspace{0.5em}
In cases where the LLM returns text outside the predefined categories and not “Other,” this typically occurs because it decides to provide its own classification as the post does not fit into any one category. These instances are considered as “Other” in Figure \ref{fig:visualizing_post_themes}.
\subsection{Prompt Template for Queries}
For query-based tasks, the following prompt template is used:

\vspace{0.5em}
\begin{lstlisting}[basicstyle=\ttfamily\footnotesize, breaklines=true, frame=single]

{pre_prompt_query}
"""
{content}
"""
{prompt_query}{instruction}
\end{lstlisting}
\vspace{0.5em}

Where:
\begin{itemize}
    \item \texttt{pre\_prompt\_query}: A pre-prompt instruction to focus on the provided context.
    \item \texttt{content}: The input content to be analyzed.
    \item \texttt{prompt\_query}: A directive to answer based on the provided context.
    \item \texttt{instruction}: The specific task or question to be addressed.
\end{itemize}

\subsection{Implementation Details}
The classification process is implemented as follows:
\begin{enumerate}
    \item The input post is passed to the language model along with the classification prompt.
    \item The model generates a response adhering to the specified format.
    \item The response is parsed to extract the category.
\end{enumerate}

\end{document}